\documentstyle[epsf,referee]{mn}

\newif\ifAMStwofonts

\newcommand{\simgt}{\lower.5ex\hbox{$\; \buildrel > \over \sim \;$}}
\newcommand{\simlt}{\lower.5ex\hbox{$\; \buildrel < \over \sim \;$}}
\ifoldfss
  \ifCUPmtlplainloaded \else
    \NewTextAlphabet{textbfit} {cmbxti10} {}
    \NewTextAlphabet{textbfss} {cmssbx10} {}
    \NewMathAlphabet{mathbfit} {cmbxti10} {} 
    \NewMathAlphabet{mathbfss} {cmssbx10} {} 
  \fi
  \ifAMStwofonts
    \ifCUPmtlplainloaded \else
      \NewSymbolFont{upmath} {eurm10}
      \NewSymbolFont{AMSa} {msam10}
      \NewMathSymbol{\upi}     {0}{upmath}{19}
      \NewMathSymbol{\umu}     {0}{upmath}{16}
      \NewMathSymbol{\upartial}{0}{upmath}{40}
      \NewMathSymbol{\leqslant}{3}{AMSa}{36}
      \NewMathSymbol{\geqslant}{3}{AMSa}{3E}

      \let\leq=\leqslant 
       
    \fi
  \fi
\fi 

\ifnfssone
  \newmathalphabet{\mathit}
  \addtoversion{normal}{\mathit}{cmr}{m}{it}
  \addtoversion{bold}{\mathit}{cmr}{bx}{it}
  \newmathalphabet{\mathbfit} 
  \addtoversion{normal}{\mathbfit}{cmr}{bx}{it}
  \addtoversion{bold}{\mathbfit}{cmr}{bx}{it}
  \newmathalphabet{\mathbfss} 
  \addtoversion{normal}{\mathbfss}{cmss}{bx}{n}
  \addtoversion{bold}{\mathbfss}{cmss}{bx}{n}
  \ifAMStwofonts
    \ifCUPmtlplainloaded \else
      \UseAMStwoboldmath
      \makeatletter
      \new@mathgroup\upmath@group
      \define@mathgroup\mv@normal\upmath@group{eur}{m}{n}
      \define@mathgroup\mv@bold\upmath@group{eur}{b}{n}
      \edef\UPM{\hexnumber\upmath@group}
      \new@mathgroup\amsa@group
      \define@mathgroup\mv@normal\amsa@group{msa}{m}{n}
      \define@mathgroup\mv@bold\amsa@group{msa}{m}{n}
      \edef\AMSa{\hexnumber\amsa@group}
      \makeatother
      \mathchardef\upi="0\UPM19
      \mathchardef\umu="0\UPM16
      \mathchardef\upartial="0\UPM40
      \mathchardef\leqslant="3\AMSa36
      \mathchardef\geqslant="3\AMSa3E

      \let\leq=\leqslant 

    \fi
  \fi
\fi 

\ifnfsstwo
  \DeclareMathAlphabet{\mathbfit}{OT1}{cmr}{bx}{it}
  \SetMathAlphabet\mathbfit{bold}{OT1}{cmr}{bx}{it}
  \DeclareMathAlphabet{\mathbfss}{OT1}{cmss}{bx}{n}
  \SetMathAlphabet\mathbfss{bold}{OT1}{cmss}{bx}{n}
  \ifAMStwofonts
    \ifCUPmtlplainloaded \else
      \DeclareSymbolFont{UPM}{U}{eur}{m}{n}
      \SetSymbolFont{UPM}{bold}{U}{eur}{b}{n}
      \DeclareSymbolFont{AMSa}{U}{msa}{m}{n}
      \DeclareMathSymbol{\upi}{0}{UPM}{"19}
      \DeclareMathSymbol{\umu}{0}{UPM}{"16}
      \DeclareMathSymbol{\upartial}{0}{UPM}{"40}
      \DeclareMathSymbol{\leqslant}{3}{AMSa}{"36}
      \DeclareMathSymbol{\geqslant}{3}{AMSa}{"3E}

      \let\leq=\leqslant 

    \fi
  \fi
\fi 

\ifCUPmtlplainloaded \else
  \ifAMStwofonts \else 
    \def\upi{\pi}
    \def\umu{\mu}
    \def\upartial{\partial}
  \fi
\fi

\title{Cosmological Constraint from the 2dF QSO Spatial Power Spectrum}
\author[K. Yamamoto]
       {K. Yamamoto\\
        Max-Planck-Institut for Astrophysics,
                Karl-Schwarzschild-Str. 1,
                D-85741 Garching, Germany
\\        Department of Physical Science, Hiroshima University,
        Higashi-hiroshima, 739-8526, Japan
}
\date{ in original form 2002 January 15
}

\pagerange{\pageref{firstpage}--\pageref{lastpage}}
\pubyear{2002}

\begin{document}
\maketitle

\label{firstpage}

\begin{abstract}
In this paper we obtain constraints on the cosmological
density parameters, $\Omega_m$ and $\Omega_b$, by comparing the preliminary
measurement of the QSO power spectrum from the
two degree field QSO redshift survey with results
from an analytic technique of power spectrum
estimation, described in this paper.
We demonstrate the validity of the analytic approach
by comparing the results with the power spectrum of
an N-body simulation.
We find a better fit to the shape of the QSO power
spectrum for low density models with $\Omega_m= 0.1-0.4$.
We show that a finite baryon fraction $\Omega_b/\Omega_m=0.2$,
consistent with observations of the CMB anisotropies and
nucleosynthesis, fits the observational result of the 2QZ
survey better,
though the constraint is not very tight.
By using the Fisher matrix technique, we investigate just how
a survey would be required before a significant constraint on
the density parameters can be made. We demonstrate that the
constraint could be significantly improved if the survey was 
four times larger.

\end{abstract}


\newpage
\section{Introduction}


Since the pioneering work of Osmer in the 1980's (Osmer 1981), QSOs
have been used to probe the high redshift Universe. It was soon shown
that QSOs were clustered (Shaver 1984; Shanks et al. 1987;
Iovino \& Shaver 1988) but results were limited due to the relatively
small QSO samples available, see Croom \& Shanks (1996) and references therein
for a summary
of early QSO clustering results. The two degree Field QSO Redshift Survey
(2QZ) will dramatically improve things as it will contain at least a factor
of 25 more QSOs than than previous QSO surveys.
The 2QZ group has recently reported preliminary clustering
results based on an initial sample of 10,000
QSOs~(Croom et~al., 2001, Hoyle et~al. 2002; 2001, Outram et~al. 2001).

Observational results and theoretical predictions are most directly
compared through the use of numerical simulations.
Indeed, the 2QZ group has utilized one of the Hubble Volume
simulations run by the Virgo consortium~(Frenk et~al. 2000) to interpret
their results.
It is a huge N-body simulation of horizon box size, containing
1 billion mass particles output along a light-cone.
However, numerical simulations of this size are slow and
expensive to run.
A simple, semi-analytic formula that reproduces numerical
results but which doesn't require intensive computation would
be useful as it allows a wide range of parameter space to be
explored in a short time frame. Fortunately, several examples appear
in the literature (e.g. Yamamoto et~al. 1999; Suto et~al. 2000; Hamana et~al.
2001b; and references therein).

In this paper, we describe such an analytic approach. We demonstrate
the validity of this technique by comparing the semi-analytic formula
with results from N-body simulations and mock 2QZ samples.
As a demonstration of the usefulness of this approach, we use the
analytic formula to place constraints on the cosmological density
parameters by comparing the theoretical predictions to the 2QZ
power spectrum of Hoyle et al.~(2002).
By using the Fisher matrix approach 
we also make predictions
for the size of the survey that would be required before the density
parameters can be tightly constrained. (see e.g., Tegmark 1997, 
Tegmark et~al. 1998)

This paper is organized as follows: In section 2, we briefly
summarize a theoretical formula for the power spectrum that
incorporates the light-cone effect, the linear redshift-space
distortions and the geometric distortion.
In section 3, we compare the analytical power spectrum formula with power
spectra results from the N-body simulation and mock QSO samples.
In section 4 we place constraints on the density parameters,
$\Omega_m$ and $\Omega_b$, by fitting the analytic formula to
the QSO power spectrum measured from the 10k 2QZ sample.
In section 5 we make predictions for how well the density parameters
may be constrained with yet larger surveys and
Section 6 is devoted to summary and conclusions.
Throughout this paper we use the unit in which the
light velocity $c$ equals $1$.

\section{Analytic Formulas}
\def\PLC{P^{\rm LC}}

In this section we briefly review an analytic formula for the spatial power
spectrum of cosmological objects. The analytic form of the power spectrum
has to include light-cone effects, redshift-space distortions
and geometric distortions. We follow, for example,
Suto et~al. (2000) and define
\begin{eqnarray}
  \PLC(k)={\displaystyle{\int dzW(z) P_0^{a}(k,z)}
  \over\displaystyle{\int dzW(z)}},
\label{PLC}
\end{eqnarray}
where
\begin{eqnarray}
  W(z)=\biggl({dN\over dz}\biggr)^2 \biggl({1 \over s^2}\biggr)
\biggl({dz\over ds}\biggr),
\end{eqnarray}
$dN/dz$ is the number density of objects per unit redshift per
unit solid angle, $s=s(z)$ is the distance-redshift relation
defined by equation (\ref{defs}),
and $P_0^{a}(k,z)$ is the local power spectrum defined on a
constant time hypersurface at redshift $z$.
Thus the light-cone effect is taken into account by averaging
the local power spectrum $P_0^{a}(k,z)$ over the redshift range.
%
The distance $s=s(z)$ is the comoving distance which, in a specially
flat universe, is given by
\begin{equation}
  s(z)={1\over H_0}\int_0^z{dz'\over
  \sqrt{\Omega_m(1+z')^3+1-\Omega_m}},
\label{defs}
\end{equation}
where $H_0=100h{\rm km/s/Mpc}$, the Hubble parameter, and
$\Omega_m$ is the total matter density parameter. In the
present paper we fix $\Omega_m=0.3$ and $h=0.7$ in equation
(\ref{defs}), consistent with the values used in the
analysis by Hoyle et~al. (2002).
\newcommand{\cpara}{c_{\scriptscriptstyle \|}}
\newcommand{\cperp}{c_{\scriptscriptstyle \bot}}
\newcommand{\qpara}{q_{\scriptscriptstyle \|}}
\newcommand{\qperp}{q_{\scriptscriptstyle \bot}}
For the local power spectrum, $P_0^{a}(k,z)$, we model
\begin{eqnarray}
  &&P_0^{a}(k,z)={1\over \cpara(z)\cperp(z)^2}
\nonumber
\\
  &&{\hspace{0.5cm}}\times\int_0^1d\mu
  P_{QSO}\Bigl(\qpara\rightarrow{k\mu\over\cpara},
  \qperp\rightarrow{k\sqrt{1-\mu^2}\over\cperp},z\Bigr),
\end{eqnarray}
where $P_{QSO}(\qpara,\qperp,z)$ is the QSO power
spectrum,  $\qpara$ and $\qperp$ are the wave number
components parallel and perpendicular to the line-of-sight
direction in the real space, and we define
\begin{eqnarray}
&&\cpara(z)={dr(z)\over ds(z)}\\
&&\cperp(z)={ r(z)\over  s(z)}
\end{eqnarray}
where $r(z)$ is the comoving distance of the universe,
found using equation (3) but with
varying $\Omega_m$.  In the present paper, we only
consider the spatially flat cosmological model with 
a cosmological constant. 
Because an analysis of the observational data is
performed by fixing the cosmological model, i.e.,
by equation (3), we need to take the geometric
distortion effect into account when comparing the
observational result with models using different
cosmological parameters.
The geometric distortion effect is included by
scaling the wave number in equation
(4) by the factors $\cpara(z)$ and $\cperp(z)$.

We model the QSO power spectrum of the distribution with
the linear distortion (Kaiser 1987) by
\begin{eqnarray}
  P_{QSO}(\qpara,\qperp,z)=
  \biggl(1+{f(z)\over b(z)}{\qpara^2\over q^2}\biggr)^2
  b(z)^2 P_{\rm mass}(q,z),
\label{PQSO}
\end{eqnarray}
where $q^2=\qpara^2+\qperp^2$, $b(z)$ is a scale independent
bias factor, $P_{\rm mass}(q,z)$ is the CDM mass power
spectrum, and we defined $f(z)=d\ln D(z)/d\ln a(z)$ with
the linear growth rate $D(z)$ and the scale factor $a(z)$.
For a phenomenological correction for the nonlinear velocity
field, a damping function might be multiplied on the right
hand side of equation (\ref{PQSO}). This is a very small
correction on the scales we consider in this work so we neglect it here.
(see e.g., Mo, Jing \& B{$\ddot {\rm o}$}rner 1997; Magira, Jing \&
Suto 2000)

The two-point correlation function, corresponding to the
spectrum (1), can be written
\begin{eqnarray}
  \xi^{\rm LC}(R)={1\over 2\pi^2}\int_0^\infty dk k^2j_0(kR)
  \PLC(k).
\label{xiLC}
\end{eqnarray}
Formula (\ref{PLC}) is obtained
under the short distance approximation, $R(=2\pi/k)\ll L$, where
$L$ is the size of a survey area. Hence the validity
of the use is limited to small length scales.
The finite size effect means we cannot obtain
a robust estimation of the correlation function or the power
spectrum on
length scales larger than the size of a survey.
For the 2QZ survey, the limit is $L\sim
10^3~h^{-1}{\rm Mpc}$.
For the correlation function (\ref{xiLC}), it was
reported that the analytic expression is in good
agreement with results of the N-body numerical
simulation (Hamana et~al. 2001a; 2001b), though the
comparison
might be limited to a finite range of length scales.

We only consider the linear power
spectrum as nonlinear effects are negligible on the length scales
on which the power spectrum from QSOs can be measured.
Hence we adopt
\begin{eqnarray}
  P_{\rm mass}(q,z)=Aq^nT(q,\Omega_m,\Omega_b,h)^2D(z)^2,
\end{eqnarray}
where $A$ is a normalization constant,
$T$ is the transfer function, and $n$ is the index of the
initial spectrum, for which we assume $n=1$ in the present
paper.
We adopt the fitting formula of the cold dark matter
transfer function by Eisenstein \& Hu (1998) which is
robust even when the baryon fraction is large.

\section{Comparison with N-body numerical simulation}
In this section we investigate how well the analytic
power spectrum reproduces a power spectrum measured from an
N-body numerical simulation and a power spectrum measured from mock
2QZ catalogues.

The simulation is one of the Hubble Volume (HV) simulations. It has a
$\Lambda$CDM cosmology with
$\Omega_{\rm m}$=0.3, $\Omega_b$=0.04 and $\Omega_{\Lambda}$=0.7, $h$=0.7
and normalized to $\sigma_8$=0.9 at
present day and with a shape parameter of $\Gamma = 0.17$.
The simulation has a light-cone
output such that the evolution of the dark matter is fully accounted for. It
extends to redshift $\approx 4$,
and covers an area of 15$\times$75deg$^2$, which we split
into three 5$\times$75deg$^2$ strips to match the geometry of the 2QZ strips.

For the first comparison, we simply take the dark matter points in real space,
over the range $0.3\leq z\leq 2.5$. We apply the selection function
of the 2QZ survey shown, for example, in Hoyle et~al. (2002), to the points
and then measure the power spectrum. We do this in each of the three
5$\times$75deg$^2$ strips and average the result together.
The analytic power spectrum
assumes the same $dN/dz$ distribution, that the points are in real space and
assumes the same cosmology. Figure 1 shows a comparison of these two power
spectra. The solid and dotted lines, respectively, show the results
of the analytic prescription and the numerical simulation.
There is good agreement (within $\sim 10$ \%) on
scales $0.01 h{\rm Mpc}^{-1}\simlt k\simlt 0.1 h{\rm Mpc}^{-1}$.

As a second comparison, we compare the analytic power spectrum to
the power spectrum of mock QSO catalogues drawn from the HV. The
mock catalogues are constructed as follows:
First, the selection function of the 2QZ is applied to the
dark matter particles in redshift-space over the redshift range
$0.3\leq z\leq 2.5$. Then the dark matter particles are biased
such that the clustering approximately matches that of the 2QZ.
It is assumed that
the QSO clustering amplitude remains constant with redshift, consistent
with the results of Croom et al. (2001), when the mock
catalogues are constructed. The points are then sparse sampled to match the
number density of the 2QZ. Again the power spectrum is measure from each
strip and the results are averaged together. Full details on the
construction of the mock catalogues are given in Hoyle (2000).

For this comparison, the amplitude of the analytic power spectrum has to
be determined. To do this, we  minimizing $\chi^2$ defined by equation
(\ref{defchi}).
\begin{eqnarray}
  \chi^2=\sum_i^N{\bigl(\PLC(k_i)-P^{\rm sim}(k_i)\bigr)^2
  \over \Delta P(k_i)^2}
\label{defchi}
\end{eqnarray}
where $P^{\rm sim}(k_i)$ is the value of the power spectrum
from the simulation at $k_i$, $\Delta P(k_i)^2$ is the
variance of errors in Figure 2, and $N$ is the number of
the data points. In the analytic modeling of the clustering
bias, we assumed the form
\begin{eqnarray}
  b(z)={b_0\over [D(z)]^p}
\end{eqnarray}
where $b_0$ and $p$ are the parameters. We determine $b_0$ to
minimize $\chi^2$ for each pair of $\Omega_m$ and $\Omega_b$.
Croom et~al (2001) report that the QSO clustering amplitude
does not show significant time-evolution, therefore
we set $p=1$. However, this choice does not
alter our conclusions. For example, when we set $p=2$
the values of $\chi^2$ change by only a few percent.
This confirms the result that the power spectrum is only
sensitive to a mean amplitude of the bias and is insensitive
to the speed of the redshift evolution
(Yamamoto \& Nishioka 2001).
Figure 2 shows that a more realistic
numerical result can still be fit by the analytic
approach even though the bias is calculated in a different way.

Figure 3 shows the $\chi^2$ contours found when different values of
the cosmological density parameters $\Omega_m-\Omega_b/\Omega_m$ are
used in the analytic formula.
Each panel corresponds to the constraint using
(a) the 17 points $(N=17)$, (b) the 16 points,
(c) the 15 points and (d) the 14 points,
respectively, of the data from left to right in Figure 2.
The cross point in each panel is the target
parameter of the simulation. In the case where we
including data points of small length scales,
the minimum of $\chi^2$ is slightly inconsistent with the
target parameter. The disagreement
on these small scales $k\simeq 0.2$ can not
be explained by including the nonlinear corrections
of the density perturbations and the finger of God
effect. It might be inferred that long mean separations
between QSOs cause this feature.
In general, however, agreement of the numerical simulation
and the analytic approach is acceptable over the range of
scales $0.01 h{\rm Mpc}^{-1}\simlt k\simlt 0.1 h{\rm Mpc}^{-1}$.

\section{Comparison with an Observational Result}
As an application, we compare the power spectrum measured from the
10k 2QZ dataset to the analytic formula as a way to constrain
cosmological parameters.
We simply define $\chi^2$ in the similar way as equation
(\ref{defchi}) but replace $P^{\rm sim}(k_i)$ with
$P^{\rm obs}(k_i)$,
i.e., the observational data, for which we use the result of
Hoyle et~al.~(2002). Figure 4 shows the observational data
points from Hoyle et~al. (2002) compared to the best fit analytic power
spectrum.
Figure 5 displays the contours of $\chi^2$ for various
cosmological models on the $\Omega_m-\Omega_b/\Omega_m$ plane.
Each panel shows the contour using the data of (a) the 17
points, (b) the 16 points, (c) the 15 points and (d) the 14
points, respectively, from left to right in Figure 4,
where the wave numbers of the observational
data used in Figure 5 are same as those in Figure 3.
{}From Figure 5 it is clear that the QSO power spectrum favors
a low density universe with $\Omega_m= 0.1-0.4$ rather than
the standard CDM model with $\Omega_m=1$.
The minimum of the $\chi^2$ is located at
$\Omega_m\simeq0.2$ and $\Omega_b/\Omega_m\simeq0.2$.
The dashed curve shows $\Omega_b=0.04$, inferred from results
of the comic microwave background (CMB) anisotropies and the big
bang nucleosynthesis (BBN) (see e.g., Turner 2001).
Our result $\Omega_m\simeq0.2$ is consistent with the previous
result of the 2QZ power spectrum, which reports the best fit
value $\Omega_mh=0.1$ (Hoyle et~al.~2002, cf. Outram et~al.~2001),
but,
is smaller than $\Omega_m\simeq0.3$, which is obtained from
the large scale structures of galaxies (e.g., Peacock et~al. 2001). 
It is unclear whether the discrepancy has any physical
meaning or not, because the $1-\sigma$ contour is broad.
An interesting fact is that the QSO power spectrum is better
explained with the finite baryonic component, though the peak
of $\chi^2$ is broad and thus the constraint is not that tight
(cf. Miller et~al. 2001, Peacock et~al. 2001).

\section{Tightness of the constraint}
\def\bfF{{\bf F}}
It is useful to investigate how much larger the 2QZ Survey
would have to be before a tight constraint on the density
parameters can be obtained. For this purpose, we adopt the Fisher matrix
approach. The Fisher matrix analysis provides a technique by which
statistical errors on parameters from a given data set can be estimated.
In general, the Fisher matrix is defined by 
\begin{eqnarray}
  \bfF_{ij}=\biggl\langle-{\partial^2 \ln L\over 
  \partial \theta_i \partial \theta_j} \biggr\rangle,
\end{eqnarray}
where $L$ is the probability distribution function of a
data set, given model parameters. Here, we follow the work of 
Tegmark (1997). Using the Fisher matrix approach, Tegmark 
investigated the accuracy with which cosmological parameters 
can be measured from galaxy surveys. Following Tegmark (1997), 
we can write the Fisher matrix element,
\begin{eqnarray}
  \bfF_{ij}={\kappa\over 4\pi^2}
  \int_{k_{\rm min}}^{k_{\rm max}} dk k^2
  {\partial P^{\rm LC}(k)\over \partial \theta_i}
  {\partial P^{\rm LC}(k)\over \partial \theta_j},
\label{defFij}
\end{eqnarray}
where we define
\begin{eqnarray}
  \kappa=\Delta\Omega\int dz W(z),
\label{defkappa}
\end{eqnarray}
where $\Delta\Omega$ is the solid angle of the survey area
and $\theta_i$ denotes the theoretical parameters. In
deriving equation (\ref{defFij}), we use the fact that
$\bar n P_{QSO}(k)\simlt O(0.1)$ where $\bar n$ is
the comoving mean number density of the QSOs.
Here we set $k_{\rm max}=0.2 h{\rm Mpc}^{-1}$
and $k_{\rm min}=0.01 h{\rm Mpc}^{-1}$.

By using the Bayse theorem, the probability distribution
in the parameter space can be written $P(\theta_i)\propto
\exp\bigl[-\sum_{ij}(\theta_i-\theta_i^{tr})
\bfF_{ij}(\theta_j-\theta_j^{tr})/2\bigr]$, where $\theta_i^{tr}$
is the target model parameters. 
Here, the Fisher matrix is the $2\times2$ matrix,
$\Omega_m$ and $\Omega_b$ matrix. We show how
accurately these two parameters can be determined for the final
2QZ sample and for a sample that is four times larger in Figure 6.
Curves in each panel indicate the $68$\%,
$95$\%, and $99.7$\% confidence region on the
$\Omega_m-\Omega_b/\Omega_m$ plane. Both panels assume
the same $dN/dz$ as the 10k 2QZ catalogue but the left panel assumes
$\Delta\Omega=750~{\rm deg.}^2$ and the right panel assumes
$\Delta\Omega=3000~{\rm deg.}^2$.
This figure shows an example of the tightness of the
constraint possible from the QSO spatial power spectrum alone.
Note that the constraint will be tighter if we combine the results with
other constraints, such as those from the
Alcock-Paczynski test (Alcock \& Paczynski 1979; Outram et al. 2001;
Hoyle et al. 2002).

\section{Summary and Conclusions}
In this paper we have shown that an analytic approach
can accurately reproduce the power spectrum of an N-body
simulation and mock QSO catalogues over the range of
length scales $k\simlt 0.1h^{-1}{\rm Mpc}$.
By performing a simple $\chi^2$ test we have also shown that
the observational QSO power spectrum is consistent with a
simply biased mass power spectrum based on the popular CDM cosmology with
a cosmological constant. We find that the analytic formula
better matches the 2QZ power spectrum if the baryon fraction
is around 20\% of the mass fraction.
The constraint is not very tight because our analysis is based
on just the initial sample of 10,000 QSOs from the 2QZ survey.
Accumulation of a number of samples will improve the
statistical errors.
Tighter constraints will be possible if we compare the results with
other cosmological constraints such as those obtained from the $\Lambda$-test.

\vspace{5mm}
{\it Acknowledgments:}~
This work was supported by fellowships for Japan
Scholar and Researcher abroad from Ministry of Education,
Science and Culture of Japan and the Deutcsher Akademischer
Austauschdienst (DAAD). The author thanks Prof. S.~D.~M.~White,
Dr. H.~J.~Mo, and the people at Max-Planck-Institute for 
Astrophysics (MPA) for their hospitality and useful 
discussions and comments.
He is grateful to the referee for useful comments on 
the earlier manuscript, which helped improve it.
He also thanks Y. Kojima for encouragement. 
The author is extremely grateful to Fiona Hoyle for 
many useful discussions and communications, and for 
providing me the results of the numerical simulation 
and the power spectrum data used in the present paper. 
This work couldn't be done without the collaborations 
with her. He acknowledge the efforts of everyone involved 
in the 2QZ project for producing a unique and valuable dataset.

\newpage
\begin{figure}
\begin{center}
    \leavevmode
    \epsfxsize=15cm
    \epsfbox[20 150 600 720]{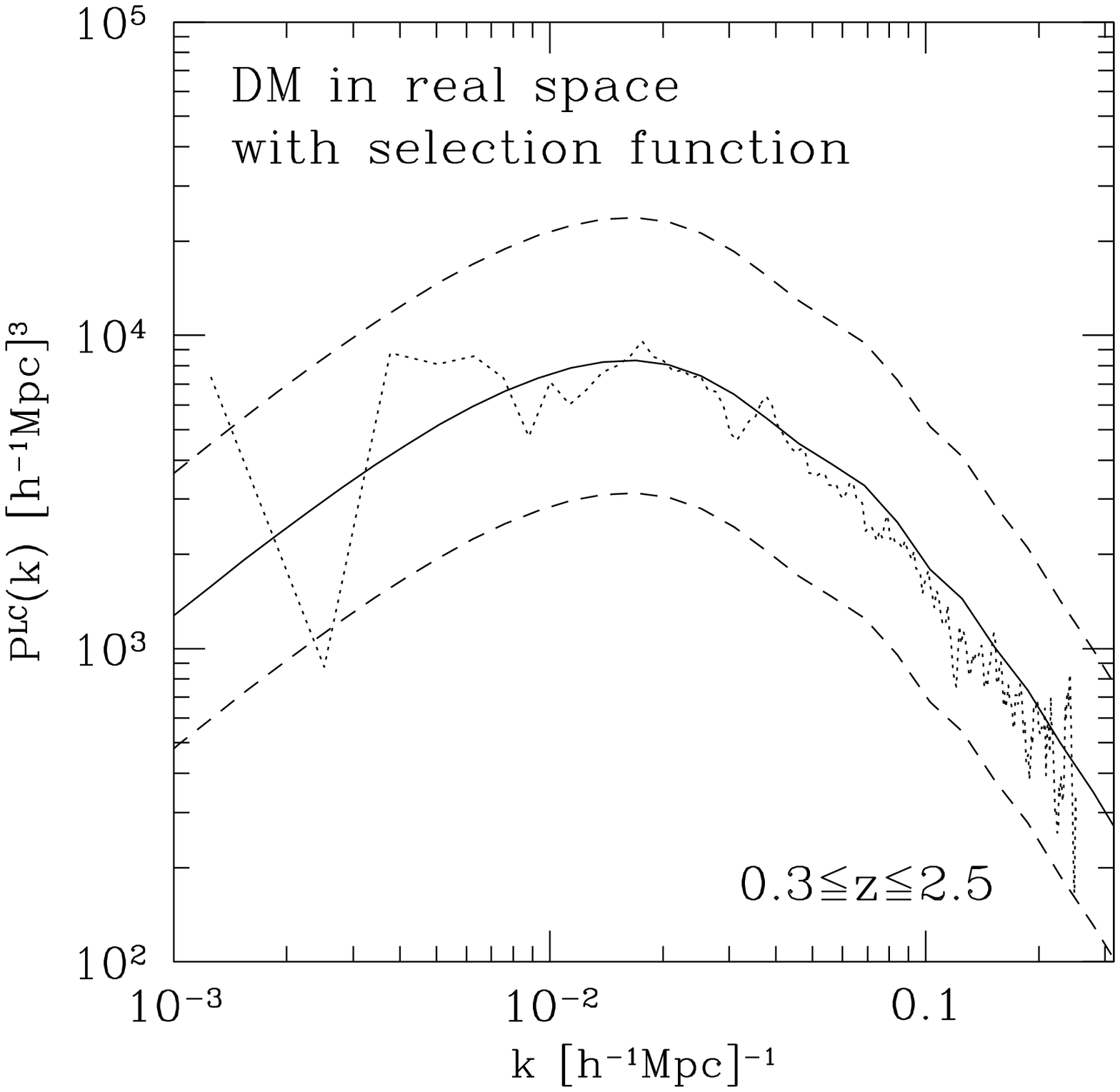}
\end{center}
\caption{
Comparison of dark matter power spectra
in real space with the 2QZ selection function applied. The dotted
curve shows the result from the numerical simulation and the solid
curve shows the result from the analytic approach.
The dashed curves are the power spectra $P^a(k,z=0)$
(the upper curve) and $P^a(k,z=2.5)$ (the lower curve).
}
\label{fiburea}
\end{figure}
\begin{figure}
\begin{center}
    \leavevmode
    \epsfxsize=15cm
    \epsfbox[20 150 600 720]{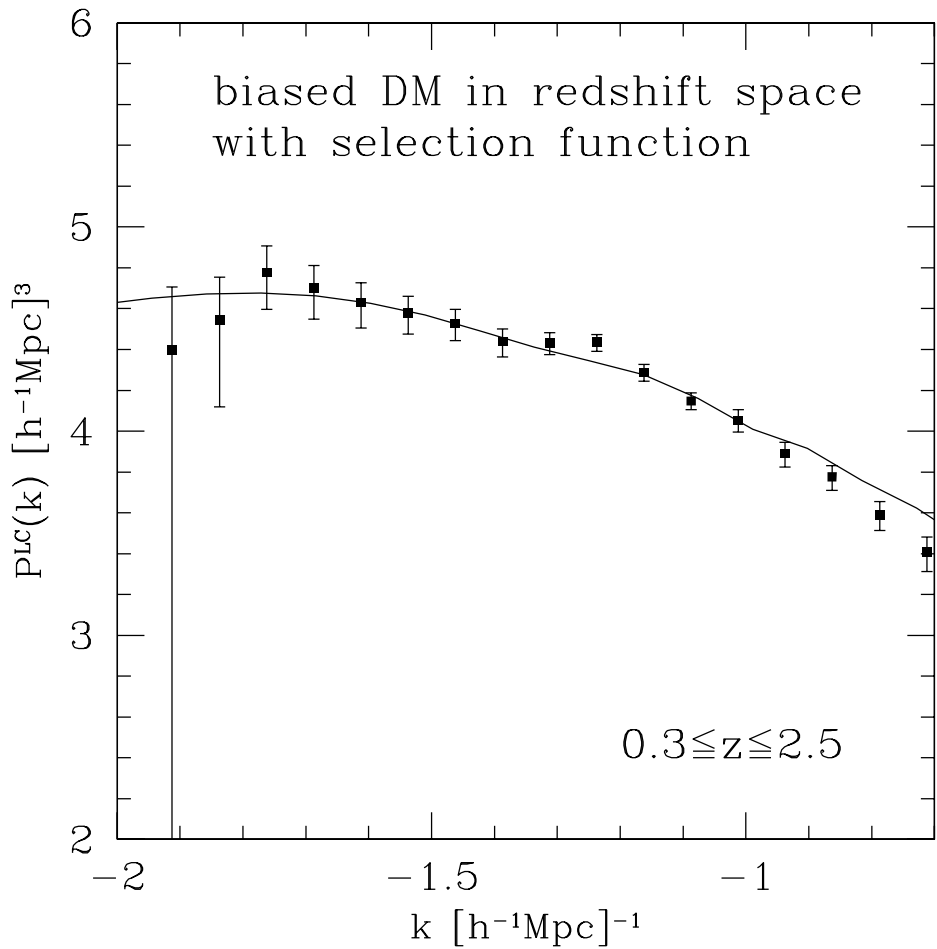}
\end{center}
\caption{
Comparison of the power spectra of biased particles
in redshift space with the 2QZ selection function applied. The points
with error bars
show the mock catalogue power spectrum from Hoyle et~al. (2002)
and the solid curve shows the result using the analytic
approach.
}
\label{figureb}
\end{figure}
\begin{figure}
\begin{center}
    \leavevmode
    \epsfxsize=15cm
    \epsfbox[20 150 600 720]{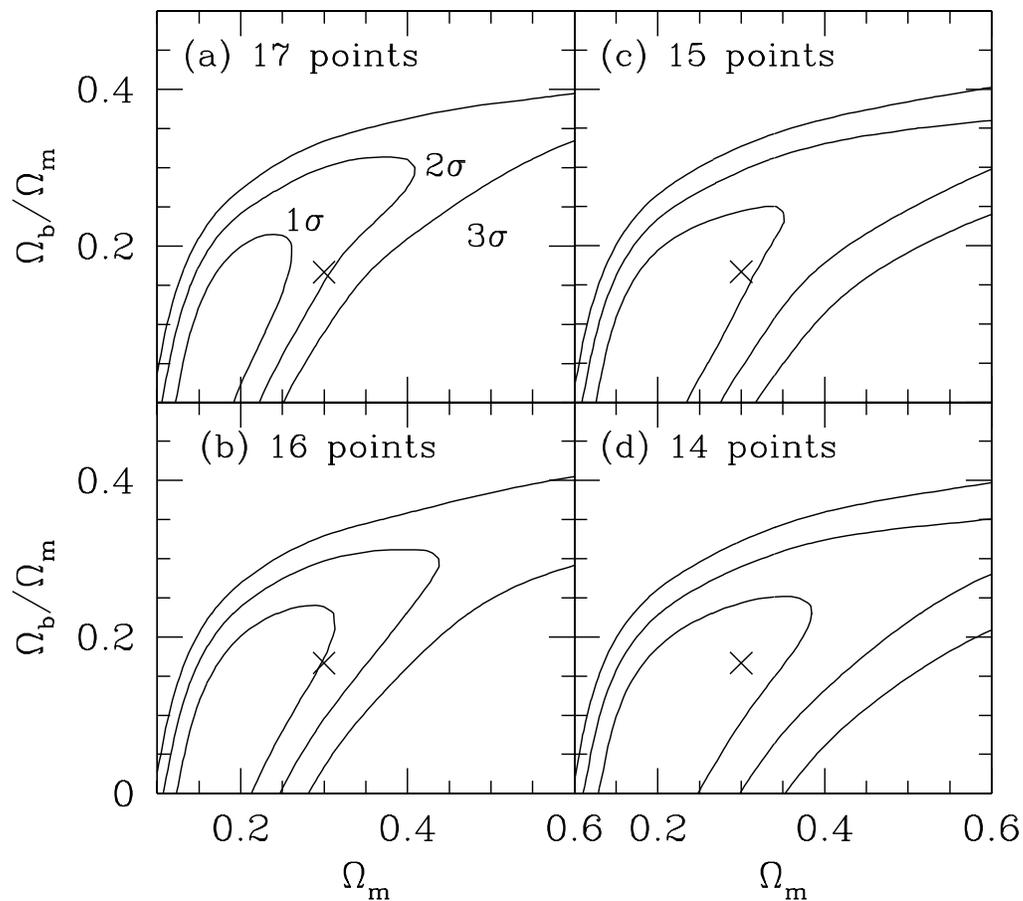}
\end{center}
\caption{
The contours of $\chi^2$ show the level to which the
power spectrum of the numerical simulation and
the analytic formula agree. The number of points used in
calculating $\chi^2$ is shown in each panel. These are
the $N$ points from the left shown in figure 2.
The cross point is the target parameter of the simulation.
}
\label{figurec}
\end{figure}
\begin{figure}
\begin{center}
    \leavevmode
    \epsfxsize=15cm
    \epsfbox[20 150 600 720]{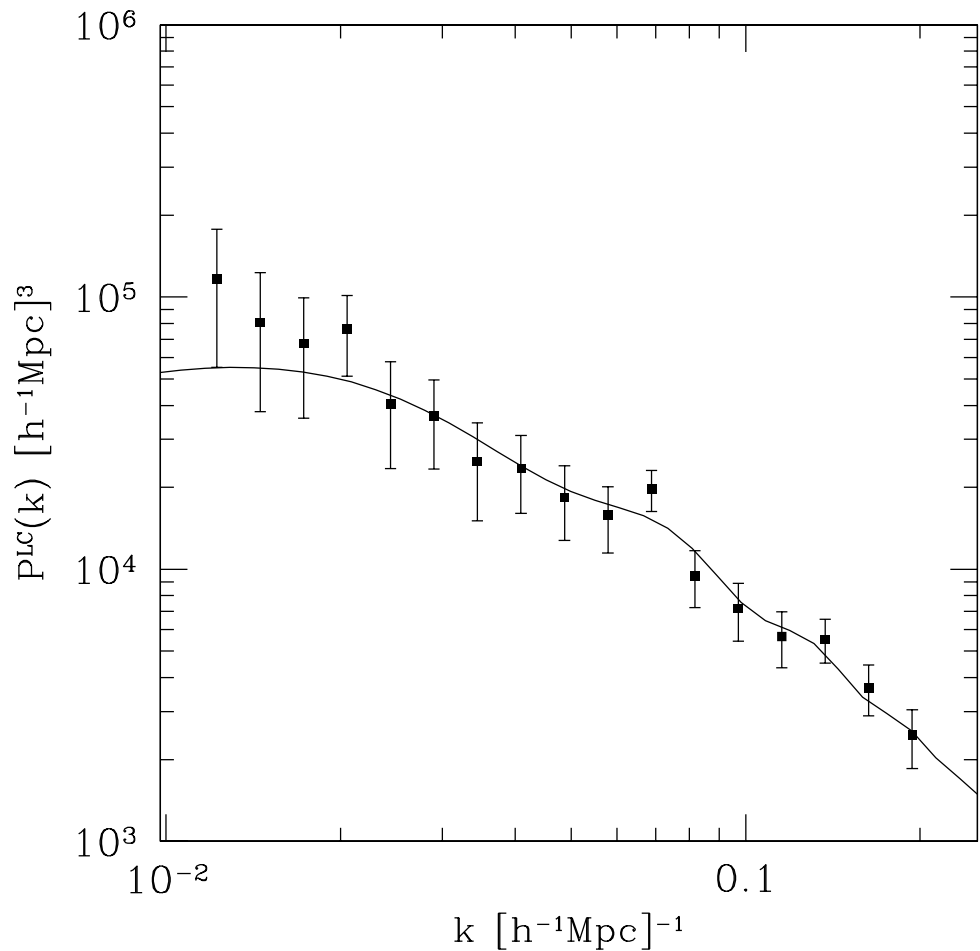}
\end{center}
\caption{
The observational data, from Hoyle et~al. (2002),
which we used in constraining the cosmological parameters.
The solid curve shows the analytic formula using the best fit
parameters $\Omega_m=0.2$ and $\Omega_b=0.04$.
}
\label{fibureb}
\end{figure}
\begin{figure}
\begin{center}
    \leavevmode
    \epsfxsize=15cm
    \epsfbox[20 150 600 720]{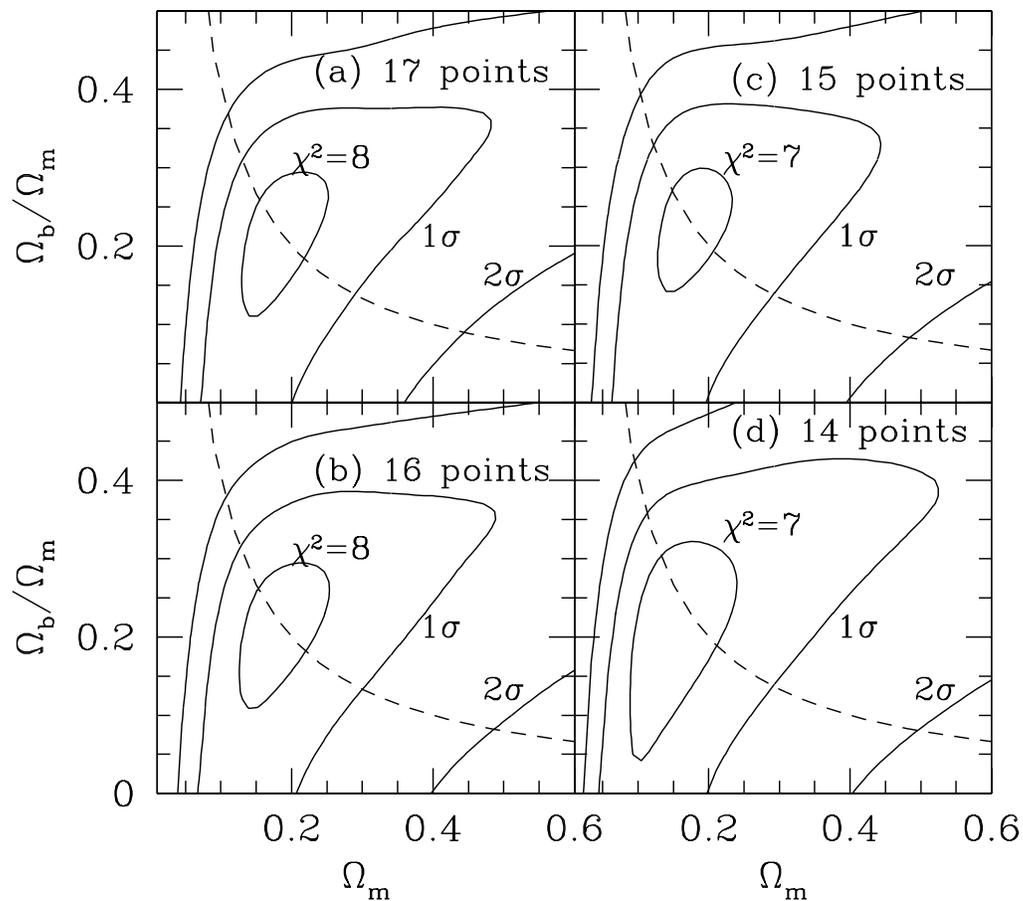}
\end{center}
\caption{
Contours of $\chi^2$ found by comparing the
observed QSO power spectrum with the analytic formula.
The format of this figure is the same as Figure 3. The number
of data points, $N$, and the wave numbers of the
observed power spectrum, $k_i$, are the same as
those in Figure 3. The dashed curve shows $\Omega_b=0.04$,
which is consistent with results from the cosmic microwave
background anisotropies and big bang nucleosynthesis.
(see e.g., Turner 2001).
}
\label{figured}
\end{figure}
\begin{figure}
\begin{center}
    \leavevmode
    \epsfxsize=15cm
    \epsfbox[20 150 600 720]{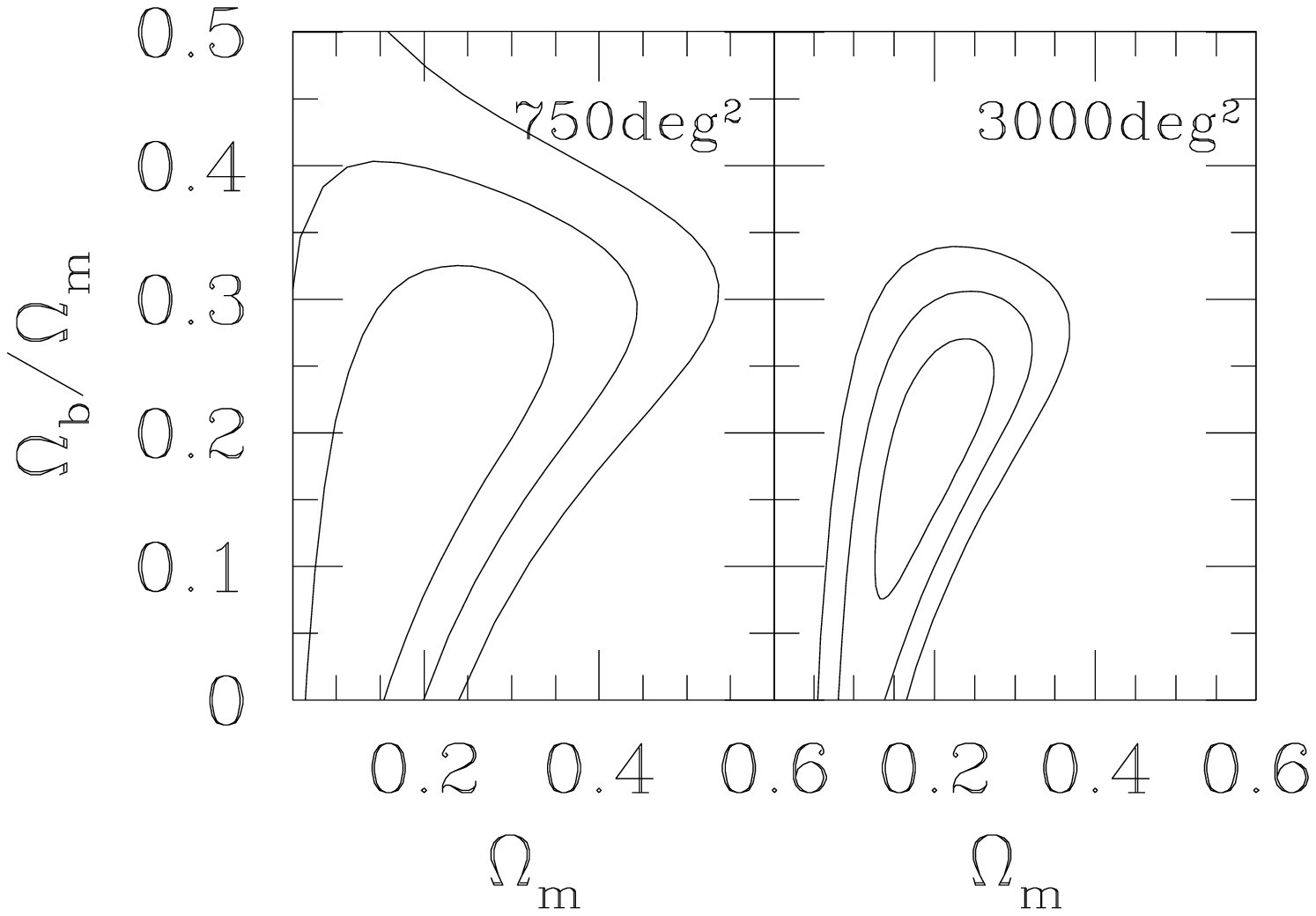}
\end{center}
\caption{
Confidence regions from data sets of QSO samples.
The curves show the $68\%$, $95\%$ and
$99.7\%$ confidence regions.
The left panel assumes a solid
angle of $\Delta\Omega=750~{\rm deg.}^2$ and the right panel assumes a solid
angle of $\Delta\Omega=3000~{\rm deg.}^2$ for the survey region.
In both cases, we assume the $dN/dz$ of the 2QZ survey, and
we set $k_{\rm max}=0.2h{\rm Mpc}^{-1}$
and $k_{\rm min}=0.01h{\rm Mpc}^{-1}$. The target parameters
are $\Omega_m=0.2$ and $\Omega_b=0.04$.
}
\label{fiburec}
\end{figure}


\newpage
\begin{thebibliography}{8.}
\addcontentsline{toc}{section}{References}

\bibitem{AP}
  Alcock C. \& Paczynski B., 1979, Nature 281 358
\bibitem{CS}
  Croom S.~M. \& Shanks T., 1996, MNRAS, 281, 893
\bibitem{yamamotoCroom01}
  Croom S.~M., Shanks T., Boyle B.~J., Smith R.~J., Miller L.,
  Loaring N. \& Hoyle F., 2001, MNRAS 325, 483
\bibitem{yamamotoEH}
   Eisenstein D.~J. \& Hu W., 1998, ApJ 496, 605
\bibitem{yamamotoFrenk}
  Frenk C.~S., Colberg J.~M., Couchman H.~M.~P., Efstathiou G.,
  Evrard A.~E., Jenkins A., MacFarland T.~J., Moore B.,
  Peacock J.~A., Pearce F.~R., Thomas P.~A., White S.~D.~M.
  \& Yoshida N., 2000, astro-ph/0007362
\bibitem{yamamotoHCS}
  Hamana T., Colombi S. \& Suto Y, 2001a, A\&A 367, 18
\bibitem{yamamotoHCSb}
  Hamana T., Yoshida N., Suto Y., Evrard A.~E., 2001b, ApJL 561, 143
\bibitem{yamamotoHoyle00}
  Hoyle F., 2000, PhD Thesis, University of Durham
\bibitem{yamamotoHoyle01}
  Hoyle F., Outram P.~J., Shanks T., Croom S.~M., Boyle B.~J.,
  Loaring N.~S., Miller L. \& Smith R.~J., 2002, MNRAS 329, 336
\bibitem{yamamotoHoylexi}
  Hoyle F., Outram P.~J., Shanks T., Boyle B.~J., Croom S.~M.
  \& Smith R.~J., 2002, accepted by MNRAS, astro-ph/0107348
\bibitem{Iovino}
  Iovino A. \& Shaver P.~A., 1988, ApJ, 330, L13
\bibitem{yamamotoKaiser}
  Kaiser N., 1987, MNRAS 227, 1
\bibitem{MagiraJS}
  Magira H, Jing Y.~P. \& Suto Y., 2000, ApJ 528, 30
\bibitem{yamamotoMillera}
  Miller C.~J., Nichol R.~C. \& Batuski D.~J., 2001
  Science 292, 2302
\bibitem{MF}
  Mo H.~J. \& Fang L.~Z., 1993, ApJ, 410, 493
\bibitem{MJB}
  Mo H.~J., Jing Y.~P. \& B{$\ddot {\rm o}$}rner G., 1997, MNRAS
  {286}, 979
\bibitem{Osmer}
  Osmer P.~S., 1981, ApJ, 247, 762
\bibitem{yamamotoOutram01}
  Outram P.~J., Hoyle F., Shanks T., Boyle B.~J., Croom S.~M.,
  Loaring N.~S., Miller L. \& Smith R.~J., 2001, 328, 805
\bibitem{yamamotoPeacock01}
  Peacock J.~A., et~al., astro-ph/0105450
\bibitem{Shanks}
  Shanks T., Fong R., Boyle B.~J. and Peterson B.~A., 1987,
  MNRAS, 296, 173
\bibitem{Shaver}
  Shaver P.~A., 1984, A\&A 136, L9
\bibitem{yamamotoSMY}
  Suto Y., Magira H. \& Yamamoto K., 2000, PASJ 52, 249
\bibitem{Tegmarka}
  Tegmark M., 1997, Phys.~Rev.~Lett. 79, 3806
\bibitem{Tegmarkb}
  Tegmark M., Hamilton A.~J.~S., Strauss M.~A., Vogeley M.~S. 
  \& Szalay A.~S., 1998 ApJ 499, 555
\bibitem{Turner}
  Turner M.~S., 2001, submitted to ApJ, astro-ph/0106035
\bibitem{yamamotoN}
  Yamamoto K. \& Nishioka H., 2001, ApJ 549, 15
\end{thebibliography}
\end{document}